\documentclass[aps,prd,groupedaddress,notitlepage,nofootinbib, noeprint]{revtex4-1}
\pdfoutput=1
\usepackage{graphicx}
\usepackage{caption}
\usepackage{subcaption}

\usepackage{amsmath,amssymb}
\usepackage[dvipsnames]{xcolor}
\usepackage{upgreek}
\usepackage[T1]{fontenc}
\usepackage{setspace}
\usepackage{footmisc}
\usepackage[normalem]{ulem}

\usepackage{textpos}
\usepackage{etoolbox}
\patchcmd{\thebibliography}{\advance\leftmargin\labelsep}
  {\labelsep=0.5cm \advance\leftmargin\labelsep}{}{}

\usepackage{epsfig}

  \newcommand{\bea}{\begin{eqnarray}}
\newcommand{\eea}{\end{eqnarray}}

\newcommand{\bi}{\begin{itemize}}
\newcommand{\ei}{\end{itemize}}

\newcommand{\be}{\begin{equation}}
\newcommand{\ee}{\end{equation}}

\newcommand{\del}{\partial}

\def\ltap{\ \raise.3ex\hbox{$<$\kern-.75em\lower1ex\hbox{$\sim$}}\ }
\def\gtap{\ \raise.3ex\hbox{$>$\kern-.75em\lower1ex\hbox{$\sim$}}\ }
\def\gl{\ \raise.5ex\hbox{$>$}\kern-.8em\lower.5ex\hbox{$<$}\ }
\def\roughly#1{\raise.3ex\hbox{$#1$\kern-.75em\lower1ex\hbox{$\sim$}}}

\newcommand{\citevain}{\cite{Vainshtein,vainintro,dyson}}

\begin{document}


\preprint{} \title{Vainshtein in the UV and a Wilsonian analysis of derivatively coupled scalars}  \author{Antonio Padilla}
\email{antonio.padilla@nottingham.ac.uk} \affiliation{School of
  Physics and Astronomy, University of Nottingham, Nottingham NG7 2RD,
  United Kingdom}
  \author{Ippocratis D. Saltas} \email{ippocratis.saltas@fzu.cz} \affiliation{CEICO, Institute of Physics of the Czech Academy of Sciences, Na Slovance 2, 182 21 Praha 8, Czechia}
  \date{\today}


\begin{abstract}
In the first part of this paper we critically examine the ultra-violet implications of theories that exhibit Vainshtein screening, taking into account  both the standard Wilsonian perspective as well as more exotic possibilities.  Aspects of this discussion draw on results from the second part of the paper in which we perform a general study of derivatively coupled scalar theories using non--perturbative exact renormalisation group techniques, which are of interest independently of their application to modified gravity. In this context, we demonstrate the suppression of quantum corrections within the Vainshtein radius and discuss the potential relation with the classicalisation conjecture. We question whether the latter can be considered a realistic candidate for UV completion of large-scale modifications of gravity on account of a dangerously low classicalisation/strong coupling scale.
\end{abstract}


\maketitle



\section{Ultra-Violet thoughts on the Vainshtein mechanism}
Phenomenologically interesting cosmological models of the early and/or late universe often invoke strongly-coupled non-linear dynamics dominated by higher-order operators. This includes {\it k}-inflation \cite{kinf}, {\it k}-essence \cite{kess}, other dark energy models \cite{edreview}, as well as a  whole slew of  large-distance modified gravity scenarios \cite{myreview, justinreview}  that rely on gravitational {\it screening} mechanisms \cite{Vainshtein,vainintro,dyson,cham1,cham2,symm1, symm2}  to evade the bounds imposed by local gravity tests \cite{will}  (see  \cite{florian}, for a recent discussion of the possible importance of screening in models of vacuum energy self-tuning and the cosmological constant problem \cite{wein, cliff, me}).
Although these setups are usually understood as effective field theories (EFTs), the strongly-coupled dynamics often pushes them beyond their naive regime of validity.  This is particularly true of cosmological models that exploit derivative self-interactions for a new gravitational degree of freedom \citevain, where strong coupling and a breakdown of perturbative unitarity often goes hand in hand with the process of Vainshtein screening \cite{power,unit}.  As a consequence, although rarely stated, most models of Vainshtein dynamics are reliant on the {\it hope} that ultra-violet effects from beyond naive cut-off are macroscopically suppressed, even though  there is no obvious reason why this should be  the case.  This is also true of generic  {\it k}-inflation scenarios\footnote{Note that there are consistent set-ups \cite{nk} that  allow one  to exploit strong coupling and yet remain below the effective field theory cut-off, essentially thanks to naive dimensional analysis \cite{georgi, manohar} and factors of $4 \pi$.}.

Following  LIGO/VIRGO's observation of neutron star merger GW170817 and its optical counterpart GRB170817A \cite{GW170817, GRB1,GRB2,GRB3,GRB4}, the space of scalar-tensor theories relevant to late Universe dynamics has been significantly reduced \cite{Spanish,Sak,vern,pedro,Amendola}. This is because gravitational waves propagate anomalously through the cosmological background even in the vicinity of heavy sources whenever the scalar mixes with curvature beyond a simple conformal coupling or  so-called kinetic gravity braiding \cite{Deffayet:2010qz}, allowing us to {\it pierce the Vainshtein screen} for those particular theories \cite{piazza}.  For those scalar-tensor theories that remain observationally viable, it has recently been suggested that the Vainshtein screen may  be pierced {inside} matter distributions \cite{marco,karim}.  Earlier attempts to made use of bounds on the time variation of Newton's constant \cite{cedric}.

Here our interest lies in understanding ultra-violet  aspects of theories with Vainshtein screening and the implications for large-scale physics.  Indeed, because higher-order operators {\it necessarily} dominate the macroscopic dynamics, the ultra-violet sector is already playing the dominant role and we  examine the extent to which we can trust the predictive power of the entire Vainshtein program applied to modified gravity. We examine this in the context of both Wilsonian and non-Wilsonian UV completions. To illustrate our immediate concerns, consider a low-energy theory described by the cubic galileon \cite{gal}, 
\begin{equation} \label{cgal}
{\cal L}=X+\frac{c}{\Lambda^3} X \square \phi,
\end{equation}
where $X=-\frac12 (\partial \phi)^2$, $c$ is some order one dimensionless number, and $\Lambda$ is the scale at which perturbative unitarity breaks down. Here we wish to focus on the dynamics of Vainshtein screening on a stable, asymptotically trivial branch of solutions, rather than on more exotic cosmological configurations\footnote{Cosmological applications of cubic galileons  have often focussed on self-accelerating configurations, which can be stable if we flip the sign of the  kinetic  term in \eqref{cgal}, but have recently been ruled out by ISW effects \cite{nogal}.}. Indeed, if we assume that the scalar couples to the trace of the energy-momentum tensor with gravitational strength, $1/M_{Pl}$, screening occurs below the so-called Vainshtein radius \cite{gal},
\be
r_V \sim \frac{1}{\Lambda} \left(\frac{M}{M_{Pl}} \right)^{1/3} \label{rv},
\ee
where $M$ is the mass of a spherically symmetric source centred at the origin. Now, it is well known that the  cubic galileon does not admit a Lorentz-invariant UV completion in the Wilsonian sense, as can be shown using analyticity arguments and the constraints they impose on low-energy scattering amplitudes \cite{nima} \footnote{These arguments rely on fundamental properties of the S-matrix, such as unitarity, analyticity (up to poles and branch cuts), crossing symmetry, and polynomial boundedness. It is independent of whether or not there are any new UV degrees of freedom.}. This has led to various speculations about the UV properties of  galileon theories \cite{class,Kovner, UVgal,apples}. What is certainly true, however,  is that one can evade the constraints from scattering amplitudes by deforming the theory with a higher-dimensional operator of  the form $g X^2/\Lambda^4$ \cite{energys, apples}. Although the coupling $g$ is technically natural on account of the fact that it breaks galileon symmetry, recent studies suggest  a new analyticity bound $g>\frac{3c^2}{10240 \pi^2} \left(\frac{E}{\Lambda} \right)^8$ \cite{Bellazzini}, where the dispersion relation is cut-off at the scale $E <\Lambda$.  This bound is accurate to order $E^2/\Lambda^2$ and we choose $E$ depending on how accurate we want the bound to be. For example, for the bound to be accurate to $1$ part in $100$ we take $E \sim 0.1 \Lambda$, yielding $g \gtrsim10^{-13}$. This should be contrasted with the fact that Vainshtein screening is spoilt by the higher-order coupling unless $g < \left(\frac{M}{M_{pl}}\right)^{-2/3}$ \cite{Y3project}.  For the Sun, this imposes the constraint $g\lesssim 10^{-26}$, which is clearly in conflict with  analyticity.   It should come as no surprise  that additional  higher-order operators can have dangerous macroscopic effects whenever the classical configuration begins to strongly probe the leading low-energy interaction. Another example of this effect can be seen in the classical solution for higher-dimensional black holes when stringy corrections are included \cite{unit}. For the Vainshtein mechanism to be trusted, at least in a Wilsonian framework, there needs to be some underlying symmetry that allows one to reliably resum {\it all} of the operators that give macroscopically important contributions within the screened environment \cite{pos}.   We are only aware of one example in which such a resummation is possible: DBIonic screening \cite{dbiV}. In this case, for gravitational strength coupling to matter,  screening kicks in at a scale $\frac{1}{\Lambda} \left(\frac{M}{M_{Pl}} \right)^{1/2}$ and one can exploit the DBI symmetry  to protect the macroscopic solution down to the strong-coupling scale $1/\Lambda \sim$ few millimetres.  As an effective theory this is acceptable although analyticity considerations again suggest that it {\it may} be challenging to embed this model in a conventional UV-complete set-up \cite{nima}. In all other phenomenologically interesting examples of Vainshtein screening, the macroscopic solution on observable scales\footnote{In a gravitational context, we define {\it observable scales} as corresponding to $0.01$-$0.1$ millimetre distances and above, as set by the experimental limit of tabletop gravity experiments \cite{will}.}  is not protected from higher-order operators (assuming order-one couplings in units of the strong-coupling scale). In these cases, for the Vainshtein mechanism to make trustworthy predictions at observable scales we must assume that the higher-order operators are suppressed, a condition that is {\it heavily} reliant on the details of the microscopic theory.  

In the next section we demonstrate how this extreme UV sensitivity  appears generically for derivatively coupled scalar theories, using the exact renormalisation group (ERG) formalism (for reviews, see \cite{Bagnuls,Berges,Gies,Polonyi,Delamotte}).  Although such theories are perturbatively non-renormalisable, our first task will be  to look beyond an effective description and  ask if there exist any asymptotically safe examples. Such a scenario would remain  predictive and under control even in the deep Vainshtein regime, thanks to the presence of an attractive non-trivial UV fixed point surrounded by a critical surface of finite dimension. However, as we will show, there are no non-trivial UV fixed points consistent with asymptotic safety within this particular class of theories. Within a standard Wilsonian framework, we conclude that our theories are at best effective theories valid up to some UV cut-off, and study the running of the couplings to low energies on generic backgrounds, generalising the results of \cite{riding}.  In particular, for fluctuations about a weakly-coupled perturbative background we find that the theory is always attracted towards the gaussian (trivial) infra-red fixed point, with the strong-coupling scale running to higher energies. In contrast, when the classical background contains gradients large enough to probe the non-linear interactions of the theory, if the Vainshtein mechanism kicks in, the running of the couplings is suppressed, with the running almost frozen to its UV-boundary condition sufficiently deep inside the Vainshtein regime. This is easily understood because Vainshtein screening serves to suppress the coupling between the quantum fluctuations and the background. {\it As a consequence, the low-energy effective theory is essentially equivalent to the UV boundary condition.} This lack of running is consistent with the existence of a strongly-coupled fixed point, but it is unrelated to the scenario of asymptotic safety.  Rather, the fixed point is infinitely strongly-coupled and we identify it with classicalisation \cite{class}, along the  lines described in \cite{Kovner}. 

Although it has been argued that classicalisation can be studied within the framework of the Wilsonian RG \cite{Kovner},  it differs considerably from more familiar Wilsonian UV completions in that it does not include the addition of new heavy degrees of freedom.  Indeed,  analyticity considerations \cite{nima} in a number of examples suggest that this may well be the correct path towards understanding the UV properties of theories exhibiting Vainshtein phenomena \cite{UVgal}.   Classicalisation postulates that scattering at ultra-high energies  remains unitary thanks to the preferential production of an ultra-large number of soft quanta, and is best motivated in a gravitational setting where the classicalons are identified as black holes. For a scalar field exhibiting Vainshtein screening  the classicalon corresponds to a macroscopic classical solution to the low-energy field equations.  Our analysis in the next section reveals a suppression of quantum fluctuations  for strongly-coupled backgrounds, consistent with expectations from classicalisation \cite{Vik, Tet4}. However, this does not mean that UV completion by classicalisation is necessarily compatible with our Universe, at least for theories that exhibit large-scale Vainshtein screening. The reason for this is entirely analogous to the question of why gravity prevents the Higgs mass from being as large as the mass of the earth (say) \cite{gia}.  In the current context of Vainshtein screening, the point is that the corresponding classicalon will contaminate scattering dynamics at colliders.  This happens because the classicalisation scale is so low, far below the electroweak scale, meaning that any sufficiently long-lived particle will become classicalized, behaving as an extended scalar configuration and decaying preferentially into large numbers of low-energy scalar quanta (see \cite{groj} for a nice discussion of this phenomena, albeit for a much higher classicalisation scale).  Note that this happens even though the coupling strengths of the would-be classicalizer are  renormalised to weaker values in the screened environment close to the earth where collider experiments are taking place. 

To get a feel for the numbers involved,  consider the cubic galileon theory described above with a strong-coupling scale $1/\Lambda \sim 1000$km consistent with massive gravity at the current Hubble scale \cite{mg}. In a screened environment close to a heavy source like the earth or the Sun, galileon fluctuations are suppressed by an environment-dependent {\it Z factor}, with $Z \sim \left(\frac{r_V}{r} \right)^{3/2}$ where $r$ is the radial distance from the centre of the heavy source and $r_V$, given by \eqref{rv}, is its corresponding Vainshtein radius \cite{energys, myreview}. This factor serves as to weaken the effective couplings so that $M^\text{eff}_{Pl} \sim M_ {Pl} \sqrt{Z} $  is  the strength of the effective coupling to matter and $\Lambda_\text{eff}\sim \Lambda \sqrt{Z}$ is  the effective strong-coupling scale \cite{energys}. Given the Vainshtein radius for the  earth, $r_V \sim 10^{17}$ m, on its surface where $r \sim 10^7$ m, we pick up a factor of $Z \sim 10^{15}$. Although the galileon fluctuations now couple to matter more than seven orders of magnitude more weakly than gravity, the strong coupling governing its self-interactions still  kicks in at a scale $1/\Lambda_\text{eff} \sim $ cm. Taking the galileon to be a classicalizer, this now sets the classicalisation scale to be sixteen orders of magnitude below the electroweak scale.  This means that sufficiently stable\footnote{Stability should be measured with respect to the classicalon formation time, which we presume to be at least of the order $1/\Lambda_\text{eff}\sim 10^{-11}$ s.}, energetic and compact configurations will now produce classicalons. To see this, consider a long-lived particle of mass $m$,  such as the proton. Even when stationary, such a particle will be described by a classicalon of size $r_* \sim \frac{1}{\Lambda_\text{eff}} \left(\frac{m}{M^\text{eff}_{Pl}} \right)^{1/3} \gg 1/m$, provided $m \gg (M^\text{eff}_{Pl} \Lambda_\text{eff}^3)^\frac14$. For the proton mass the classicalon size corresponds to around $10$ nanometers, eight orders of magnitude larger than the Compton wavelength, greatly enhancing its scattering cross-section. Furthermore, classicalon dynamics is generically dominated by decay into a large number of light scalar quanta \cite{class, groj}, which could yield a significant amount of missing energy in colliders and other possible signatures (see \cite{Higgsplprec} for precision measurements in the closely related Higgsplosion scenario \cite{Higgspl1, Higgspl2, Higgspl3}).  Of course, the discussion here is somewhat heuristic. It would be very interesting to perform a more thorough analysis of the constraints colliders place on classicalisation in this context. We emphasize that our concerns relate to the low scale of classicalisation required in applications to theories with Vainshtein screening, and not to classicalisation in general.

Returning to the standard Wilsonian perspective, how should we interpret the lack  of running for effective theories when Vainshtein screening is at work? In \cite{riding}, in the context of Vainshtein screening for $P(X)$ theories \cite{PX}, the authors focussed on the fact that the low-energy theory would essentially be insensitive to the Wilson cut-off, consistent with a natural framework. Whilst this is certainly true, we also note that the UV boundary condition, in the absence of an attractive, non--trivial UV fixed point, is controlled by its matching to whatever new Physics completes the theory. When running is suppressed, the low-energy effective theory and all of its predictions become {\it extremely} sensitive to the details of the UV matching. This is both a concern and an opportunity.  A concern because the power of effective field theory is lost and our ability to make predictions is no longer guaranteed to be insulated from the full tower of irrelevant operators that could be present. An opportunity because if Nature has indeed chosen this path, then we have an observational window on macroscopic scales that peers deep into the ultra-violet sector of the theory.  

In our view, the Vainshtein mechanism is only viable when there is a known symmetry that allows one to resum an infinite tower of large operators, as in DBI, or when it occurs within a UV complete framework.  Although there is no known example of a standard Wilsonian UV completion of Vainshtein screening, we could imagine one built from a combination of light and heavy fields for which the light fields exhibit some sort of macroscopic screening close to a source thanks to a non-perturbative  mixing between the light and heavy sectors. However, once this happens, there is no sense in which we can regard the heavy sector as being decoupled and there is absolutely no reason to expect the screening effects to be well-approximated by a truncated low-energy effective Lagrangian, as is normally assumed to be the case. For a non-standard UV completion via classicalisation, we expect that the classicalisation scale needed for a phenomenologically  interesting modified gravity set-up is far too low and anticipate problems for collider physics. More work to investigate this latter possibility is certainly required.

\section{The Exact Renormalisation Group and Derivatively Coupled Scalars}

Motivated by our interest in the UV properties of theories that exhibit Vainshtein screening, we now investigate the behaviour of derivatively coupled scalars using  the exact renormalisation group formalism.  Our discussion builds on other attempts to investigate quantum corrections in similar  theories:  in particular, loop corrections to higher-derivative scalar theories have been discussed in \cite{Shap,Tet1,Tet2}, with the exact renormalization group employed in \cite{riding,Tet3} \footnote{For applications of the exact renormalization group in other scalar-field theory setups with derivative interactions see also  \cite{Morris:1997xj, Safari:2017tgs,Rosten}.}. The suppression of quantum fluctuations in a similar context has been highlighted in \cite{riding, Brax}, as well as in \cite{Vik,Tet4}, the latter from the perspective of classicalisation \cite{class}. Galileon non-renormalisation theorems, including the extensions to heavy loops,  have been proven in \cite{Kurt1,Kurt2} and will be  perfectly consistent with our analysis.  Quantum corrections to galileons using a covariant framework were computed in \cite{Saltas1,Saltas2}.

The ERG provides a powerful framework for computing  the running of the effective action at all orders in loops, implementing the Wilsonian idea of integrating out fluctuations, and facilitating the search for  non--perturbative RG fixed points. Formally, a solution to the  ERG equations corresponds to a complete quantum solution to a given   field theory. In practise, however, the equations are far too complicated to solve exactly, and one typically employs various approximation methods.  To derive the ERG equation, we modify the Euclidean path integral  by introducing an infrared regulator, 
\be
e^{W_k[J]} = \int \mathcal{D}\phi \, e^{- S[\phi] +  J\cdot \phi - \frac{1}{2}\phi \cdot R_{k} \cdot \phi},
\ee
where $J(x)$ corresponds to an external source for the field $\phi(x)$ and $R_{k}$ is an appropriately chosen regulator function built out of a sliding RG scale $k$, implementing IR regularisation. Taking the Legendre transform of the generator of connected Green's functions, $W_k[J]$, yields the scale-dependent  effective action, $\Gamma_k[\phi]$, where $\phi=\frac{\delta W_k}{\delta J(x)}$ is the average classical field. $\Gamma_k[\phi]$ qualitatively captures the effects of integrating  out  all quantum fluctuations above the sliding RG scale, $k$, up to the cut-off, and can be shown to satisfy the ERG equation \cite{Wetterich,Morris}
\begin{align}
\partial_t \Gamma_k = \frac{1}{2} \text{Tr} \left[ \frac{\partial_t R_k}{\Gamma_k^{(2)} + R_k} \right]. \label{ERGE}
\end{align}
Here, the supertrace ''$\text{Tr}$" sums over all momenta, spacetime and internal indices,  
$
 \Gamma_k^{(2)} \equiv \frac{\delta^{(2)}\Gamma_k}{\delta \phi(x) \delta \phi(y)}
$
denotes the inverse propagator, and we define the RG-time derivative $\partial_t \equiv k \partial_k$. The functional form of $R_k$ needs to be chosen so that the integral is regular in both the UV and the IR, and has the effect of suppressing quantum fluctuations below the scale $k$. Accordingly, for a theory cut-off at some UV scale $\Lambda_0$, $\Gamma_{k \to \Lambda_0}$ is identified  as the bare action, while $\Gamma_{k \to 0}$ is the usual effective action with {\it all }quantum fluctuations integrated out.  Note that  equation \eqref{ERGE} is simply the Legendre transform of Polchinski's functional RG equation \cite{Polchinski}. For further details on the ERG formalism, and asymptotic safety,  we refer the reader to the following review articles  \cite{Bagnuls,Berges,Gies,Polonyi,Delamotte,Rosten,Percacci,Niedermaier}.

As stated above, our interest here lies in derivatively coupled scalar theories whose strongly-coupled non-linear dynamics have been applied to cosmology, both at early and at late times (see eg \cite{kinf,kess,myreview,justinreview,edreview}). Of course, generically  these theories are perturbatively non-renormalisable, and receive quantum corrections in the form of a tower of higher-order operators.  We will focus on theories of a single scalar field, invariant under constant shifts, consistent with a Goldstone boson of some spontaneously broken underlying symmetry. Generically, the corresponding effective action will take the form 
$
\Gamma[\phi]=\int d^4 x P(\del \phi, \del \del \phi, \ldots)
$
where we suppress the tensor indices in the derivatives, noting only that the form of each individual operator respects Lorentz invariance. The presence of higher derivatives indicates the existence of new heavy degrees of freedom in the theory typically introduced to preserve unitarity at high energies (for a nice discussion, see \cite{call}).
In a standard Wilsonian set-up, the Wilson coefficients at the UV boundary of our RG flow are set by proper matching  on to the UV theory with the heavy degrees of freedom included. 

In practise, such a general ansatz for the effective action is intractable from an ERG perspective. For this reason, we truncate to theories of the form 
\be
\Gamma[\phi]=\int d^4 x P(X, B),
\ee
where $X=-\frac12 (\partial \phi)^2$ and $B=\square \phi$.  Provided $P$ is non-linear in $B$, this truncation still extends beyond the so-called Horndeski class of second order systems \cite{horn,dgsz}, thereby allowing us to probe the quantum effects of any additional heavy degrees of freedom in our theory.  For the scale-dependent effective action, each operator is equipped with a scale-dependent coupling. In particular, we will assume a polynomial ansatz,  $\Gamma_k[\phi]=\int d^4 x P_k(X, B)$, where $P_k(X, B) =\sum_{n,m \geq 0} c_{n, m}(k)
 X^n B^m$.  At the ``true" cut-off of the theory, $\Lambda_0$, the Wilson coefficients are assumed to be order one in units of a (possibly different) scale $\Lambda$ inherited from the UV theory, that is  $c_{n, m}(\Lambda_0) \sim{\cal O}(1)/\Lambda^{4n+3m-4}$.  For the bare action to respect perturbative unitarity  we require $\Lambda > \Lambda_0$. However, we are also  interested in effects at strong coupling, so we will also allow for $\Lambda \lesssim \Lambda_0$, with $\Lambda_0$ possibly even being infinite. As a caveat we note that the polynomial ansatz is not especially well justified at strong coupling, although it is consistent with most of the phenomenological models one finds in the literature. Furthermore, whenever we push our theory beyond the naive unitarity cut-off, we are implicitly assuming that any higher derivatives do not yield problems with Ostrogradski ghosts \cite{ostro1,ostro2} (see, eg, \cite{bhorn}).
 
  We now evaluate the ERG equation \eqref{ERGE} in the presence of a generic background field $\phi$.  To compute the inverse propagator, we simply evaluate the scale-dependent field equations, perturb them to leading-order in the field, $\phi \to \phi+\delta \phi$, and extract the corresponding operator according to the relation
 \be
 \delta \left[ \frac{\delta \Gamma_k}{\delta \phi(x)}\right]=\int d^4 y  \frac{\delta^{(2)}\Gamma_k}{\delta \phi(x) \delta \phi(y)}  \delta \phi(y).
 \ee
 This yields the kinetic operator
 \be
 \int d^4 y  \frac{\delta^{(2)}\Gamma_k}{\delta \phi(x) \delta \phi(y)}  \delta \phi(y)=\left(
 Z_\mu \del^\mu+Z_{\mu\nu} \partial^\mu \partial^\nu  + 2\del_\mu P_{BB} \square \del^\mu +P_{BB}  \Box^2  \right)\delta \phi(x),
 \ee
 where
 \bea
 Z_\mu &=&\del_\mu P_X-\del^\nu(P_{XX} \del_\mu \phi \del_\nu \phi)-\square (P_{XB} \del_\mu \phi), \nonumber \\
 Z_{\mu \nu} &=& (P_X + \partial^{\alpha}(P_{XB}\del_\alpha \phi)+\square P_{BB})\delta_{\mu \nu}- P_{XX} \del_\mu \phi \del_\nu \phi+  \delta_{\mu \nu}  - 2 \partial_\mu(P_{XB} \del_\nu \phi), \label{def: M}
 \eea
 and we have dropped the index $k$ for brevity.

The ERG equation makes use of  a regularisation scheme in order to implement the integrating out of fast modes, in the Wilsonian sense. To this end, we choose a generalised version of   the so--called optimised cut--off \cite{Litim,Litim2}, given by
\begin{align}
R_{k} =\left[ -Z_1(k)(k^2 + \Box)+Z_2(k)(k^4-\square^2) \right]\Theta(k^2 + \Box),
\end{align}
where $k$ is the sliding RG scale, implementing IR regularisation. $Z_1(k) \equiv c_{1, 0}(k)$ is the (scale--dependent) coupling of the canonical term in the effective action, ie the usual  wave--function renormalisation for the scalar, while $Z_2(k)\equiv 2c_{0,2}(k)$ is the higher-derivative two-point vertex on a trivial background.   Expanding the fluctuations in terms of plane waves, $\delta \phi(x) =\int \frac{d^4 p}{(2\pi)^4 }\delta \phi(p)e^{ipx}$,  allows us to pass to Fourier space in the usual way, thereby trading  $\partial^\mu \to i p^{\mu}$.  We also note that the regulator is constructed in such a way such that it controls the spectrum of the kinetic operator as
$
\Gamma^{(2)} \to \Gamma^{(2)} + R_{k}.$ Indeed, on a trivial background low-momentum modes with $p<k$ no longer fluctuate since  $\Gamma^{(2)} + R_{k} \to $ constant and non-zero, consistent with the idea that only the high-momentum fluctuations are integrated out at finite coupling.

 To proceed, we make the following ansatz for the background, $\phi=\frac{B}{8} x_\mu x^\mu $ for constant $B=\square \phi$.  This choice ensures that the ERG equation \eqref{ERGE} remains closed   under the $P(X, B)$ truncation, yet it is sufficiently general for us to treat $X=-\frac{B^2}{32} x_\mu x^\mu$ and $B$ as independent.  The philosophy is analogous to the local potential approximation employed in generic scalar field theories \cite{LPA,LPAMorris}, or indeed the constant curvature background used in $f(R)$ truncations in the ERG approach to quantum gravity \cite{LPAQG,Falls:2016msz}. In any event, it will be sufficient for us to elucidate the interesting physics.
 
 With this ansatz for the background,  it is straightforward to show that the ERG equation \eqref{ERGE} yields the following,
 \be \label{erg1}
 \del_t P=\frac12 \int_{|p| \leq k} \frac{d^4 p}{(2\pi)^4} \frac{\del_t(Z_2 k^4-Z_1 k^2)+p^2\del_t Z_1-p^4 \del_t Z_2}{\alpha_0 + i\alpha_1  x_\mu p^\mu + \alpha_2  ( x_\mu p^\mu)^2}, 
 \ee
 where 
 \bea
 \alpha_0 &=& -Z_1 k^2+Z_2 k^4+p^2\left( Z_1-P_X-\frac{B}{2} P_{XB} -\frac{XB}{2} P_{X^2 B}+\frac{B^2}{4} P_{XB^2}+\frac{XB^2}{8} P_{X^2 B^2} \right)+p^4 \left( -Z_2+P_{BB} \right), \nonumber \\
 \alpha_1 &=& -\frac{B^2}{32} \left( 12 P_{XX}  +4XP_{X^3} -3B P_{X^2 B}-XB P_{X^3 B}\right) +p^2\left( \frac{B^2}{8} P_{X B^2} \right), \nonumber \\
 \alpha_2 &=& \frac{B^2}{32} \left(  2 P_{XX}-B P_{X^2 B} \right),
 \eea
and we remind the reader that we have dropped the running index $k$ on $P(X, B)$ in the interests of brevity.  Notice also  the finite range of integration, resulting from the regulator in the numerator of the ERG equation \eqref{ERGE}. To perform the integration, we assume that $x^\mu$ aligns with a fixed axis that makes an angle $\theta$ with the Euclidean momentum.  This allows us to write $x_\mu p^\mu= x p \cos \theta$, where $x=\sqrt{x_\mu x^\mu}=\sqrt{\frac{-32X}{B^2}}$ and $p=|p|=\sqrt{p_\mu p^\mu}$. Thanks to the cylindrical symmetry, we can also perform two of the angular integrations so that, $$\int_{|p| \leq k} {d^4 p} \to 4 \pi \int_0^k dp p^3 \int_0^\pi d\theta \sin^2\theta\ .$$ The resulting RG flow  is a complicated partial differential equation that is functionally dependent on   $X$, $B$ and the RG scale $k$. 

With the RG flow at hand,  we proceed by asking whether such theories can be asymptotically safe, and therefore UV complete. To this end, we look for a non-gaussian (non-trivial) fixed point at finite dimensionless couplings\footnote{In order to describe the fixed point independently of the RG scale, 
it is often convenient to pass to dimensionless variables $x^\mu \to \tilde x^\mu/k, \phi \to k  \tilde \phi$,  such that $d^4 x  ~\frac{c_{n, m}(k)}{\Lambda^{4n+3m-4}}
 X^n B^m \to  d^4 \tilde x~\tilde c_{n, m}(k)
 \tilde X^n \tilde B^m$
where we have the dimensionless couplings given by \eqref{rescale}. }, 
 \be \label{rescale}
 \tilde c_{n, m}(k) =c_{n, m}(k) k^{4n+3m-4}.
 \ee 
At the fixed point, the non-vanishing $\tilde  c_{n, m}(k) $ should be independent of the sliding RG scale $k$. In the appendix, we prove   the absence of such a  fixed point within the class of $P(X)$ theories. Note that although this is proven for the particular background ansatz employed here, it is clear that its validity extends to more general backgrounds\footnote{In contrast, proving the existence (rather than absence) of a fixed point for a particular background ansatz does not guarantee that the fixed point will exist for a more general ansatz.}.  To go beyond $P(X)$, we must be content to prove the absence of the fixed point using a truncated polynomial expansion, working up to operators of some particular dimension (see eg \cite{Morris1,Morris2}). Working up  to dimension eleven operators, we  approximate the running of  $P(X, B)$ on perturbative backgrounds as follows
\be \label{trunc}
P_k(X, B) \approx V(k)  +Z_1(k) X+c_{0,2}(k) B^2+ c_{1,1}(k) XB+c_{2,0}(k) X^2+ c_{0,3}(k) B^3+ c_{1,2}(k) XB^2+ c_{2,1}(k) X^2 B. 
\ee
Switching to dimensionless couplings, we equate operator coefficients on either side of the ERG equation \eqref{ERGE} and arrive at a set of flow equations  of the form $\del_t \tilde c_{n, m}=\beta_{n, m}(\tilde c_{i, j})$. The corresponding $\beta$-functions  are given by lengthy formulae that are not especially illuminating. However, we do note that the only fixed point in the reals corresponds to the trivial one, with $\tilde V=\frac{1}{128 \pi^2}, Z_1\neq 0$ and $\tilde c_{n, m}=0$. Further, linearisation of the beta functions about this fixed point yields, 
\bea
\beta_{{V}} &\approx&-4\delta \tilde V -{\frac {192\,{\pi }^{2}Z_{{1}} \tilde c_{{0,2}}-2\,\tilde c_{{1,2}}+\tilde c_{{2,0}}}{3072\,{
\pi }^{4}{Z_{{1}}}^{2}}},
 \\
\beta_{{Z}} &\approx& \frac{1}{32}\,{\frac {\tilde c_{{1,2}}-2\,\tilde c_{{2,0}}}{{\pi }^{2}Z_{{1}}}} ,\\
\beta_{{0,2}}&\approx& 2\tilde c_{0,2}-\frac{1}{32}\,{\frac {\tilde c_{{1,2}}}{{\pi }^{2}Z_{{
1}}}} ,
 \\
\beta_{{1,1}}&\approx& 3  \tilde c_{{1,1}} -\frac{3}{32}\,{\frac {\tilde c_{{2,1}}}{{\pi }^{2}Z_{{
1}}}}  ,
\eea
and $\beta_{n, m} \approx (4n+3m-4) \tilde c_{n, m}$ otherwise. The associated eigenvalues are given by $\{-4,0,2,3,4,5,6,7\}$, with the marginal direction corresponding to deformations of the wavefunction $\delta Z_1$. The only attractive direction in the UV is along deformations of the vacuum energy $\delta \tilde V$, which is, anyway, non-dynamical in the absence of gravity.  The rest of the couplings are repelled away from the fixed point  in the UV.   At least at the level of our truncation we conclude that there is no  fixed point compatible with an asymptotically safe interacting theory.  Instead, we find that all interactions are attracted towards their trivial value in the infra-red. 

An interesting check of our work is seen in the limit of galileon invariance \cite{gal}, $\phi\to \phi+a_\mu x^\mu+b$, where the symmetry reduces the  truncation down to 
$$
P_k(X, B) \approx  V(k)  +Z_1(k) X+c_{0,2}(k)B^2+ c_{1,1}(k) XB+ c_{0,3}(k) B^3
\ .$$   Of course, the resulting  flow preserves the galileon symmetry, but more importantly we find that $\beta_{1, 1}=3 \tilde c_{{1,1}}$ {\it identically}, without any further approximation. This ensures that the cubic galileon coupling $c_{1,1}$ does not run,  even in the presence of heavy states arising from the higher derivatives in our action, and reflects the enhanced galileon non-renormalisation theorem discussed in \cite{Kurt2}. In contrast, we find that there {\it is} running of the  other couplings, $c_{0,2}, ~c_{0,3}$, corresponding to interactions with higher powers of $B$.

\begin{figure}
\centering
\begin{subfigure}{.3\textwidth}
  \centering
    \caption{$\Lambda/\Lambda_0=10$}
 \includegraphics[width=.9\linewidth]{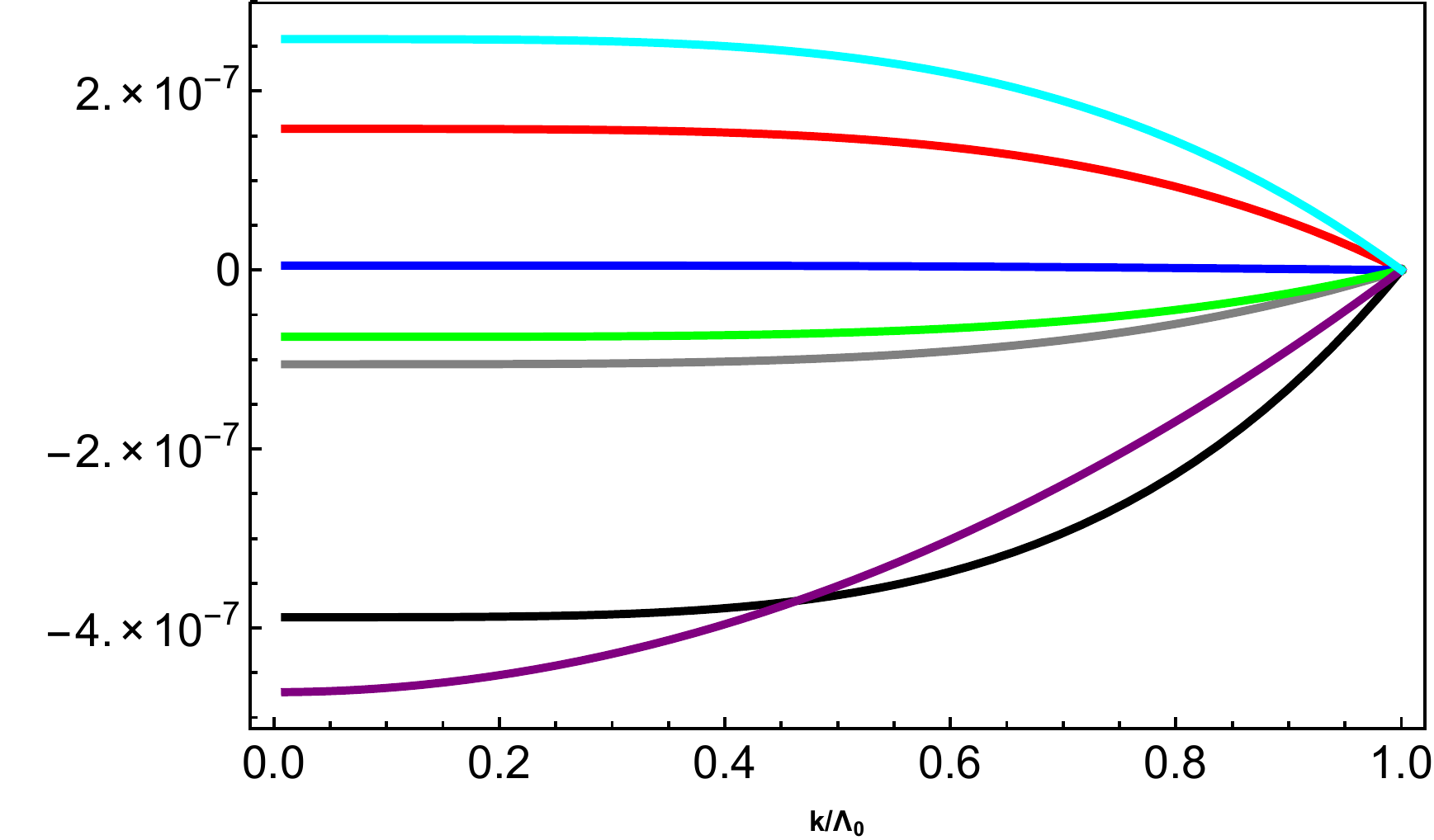}
  \label{fig:sub1}
\end{subfigure}
\begin{subfigure}{.3\textwidth}
  \centering
    \caption{$\Lambda/\Lambda_0 =1$}
  \includegraphics[width=.9\linewidth]{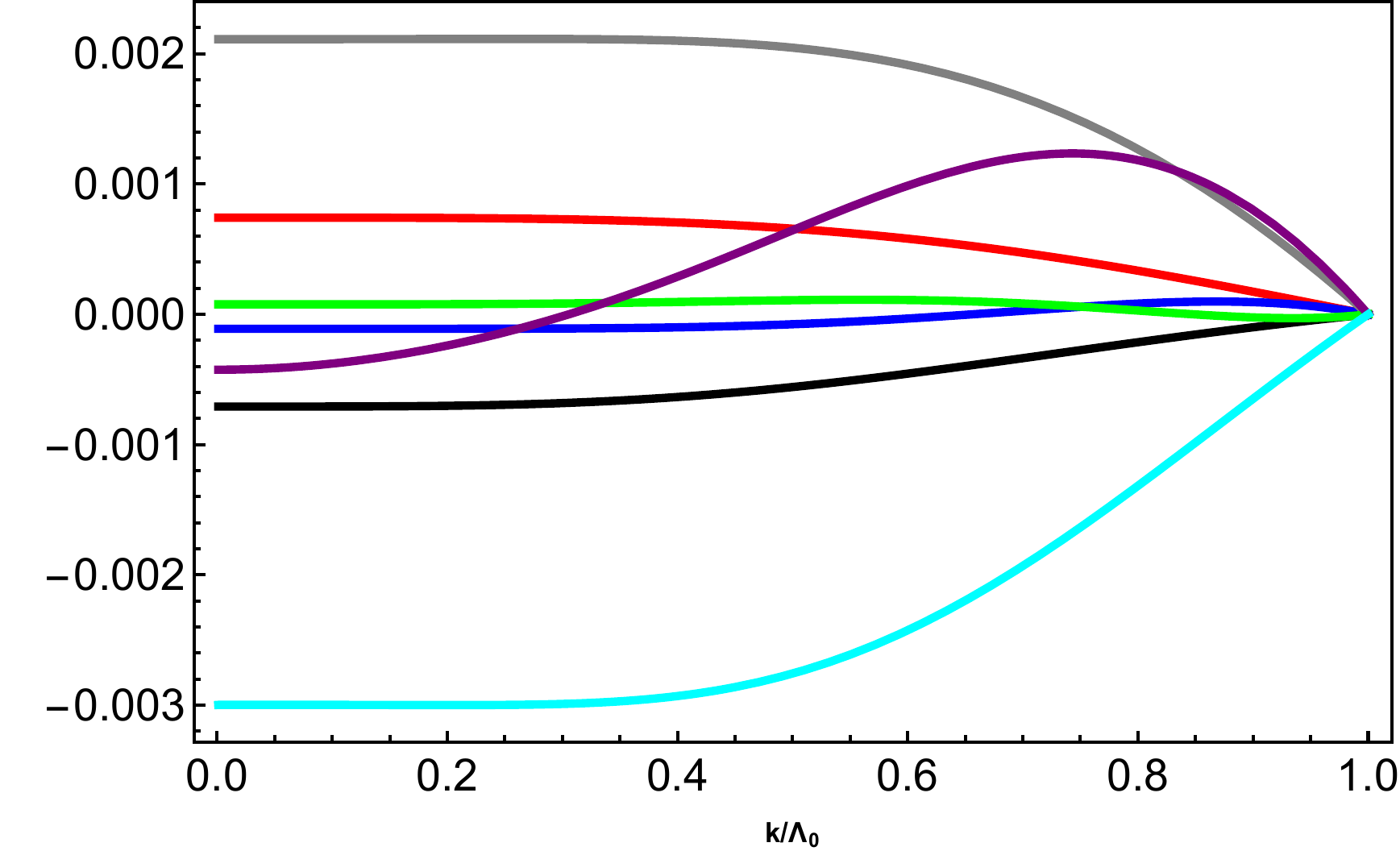}
  \label{fig:sub2}
\end{subfigure}
\begin{subfigure}{.3\textwidth}
  \centering
    \caption{$\Lambda/\Lambda_0=0.1$}
  \includegraphics[width=.9\linewidth]{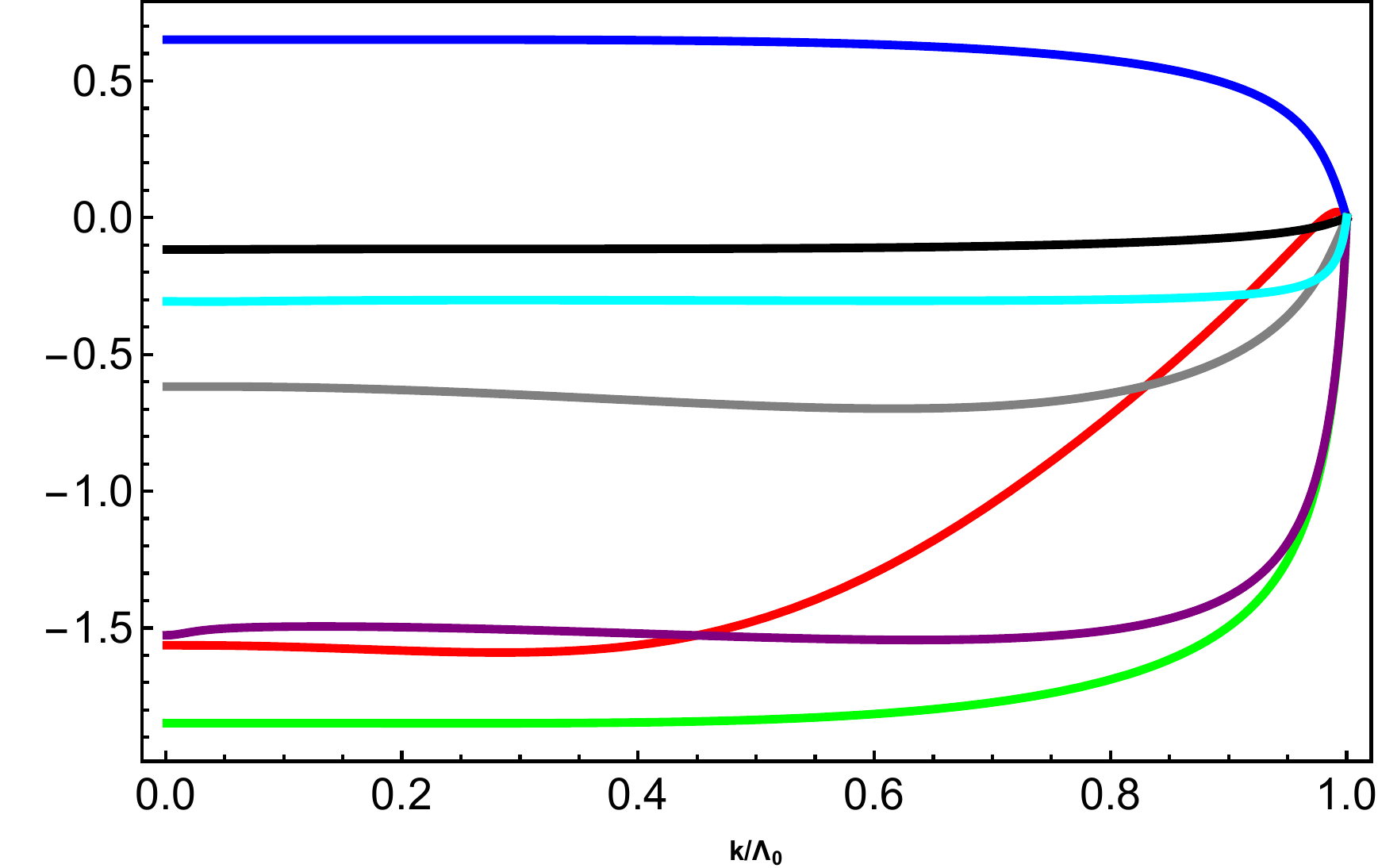}
  \label{fig:sub3}
\end{subfigure}%
\caption{
RG flows for the truncation (\ref{trunc}) after solving the ERG equation (\ref{erg1}), displaying the running of the following dimensionful couplings:  $\ln\left(\frac{Z_1(k)}{Z_1(\Lambda_0)}\right)$ (red), and 
$\ln \left(\frac{c_{n,m}(k)}{c_{n,m}(\Lambda_0)}\right)$ with $c_{n,m} = $ $c_{0,2}$ (blue), $c_{1,1}$ (yellow), $c_{2,0}$ (black), $c_{0,3}$ (green), $c_{1,2}$ (purple), $c_{2,1}$ (cyan), with initial data $k_{\text{initial}} = \Lambda_0$, $Z_1(\Lambda_0)=1$, $c_{0,2}(\Lambda_0)=-1/\Lambda^{2}$, and $c_{n, m}(\Lambda_0)=1/\Lambda^{4n+3m-4}$ otherwise.  Note that we chose to take $c_{0,2}(\Lambda_0)<0$ to avoid running into a heavy pole in the propagator, while in case (a) the curve of $c_{1,2}$ is rescaled by an overall factor of $10^{-2}$ for illustration purposes. $\Lambda$ and $\Lambda_0$ denote the bare strong-coupling scale and the theory's cut-off respectively. Case (a) corresponds to weakly-coupled initial conditions while cases (b) and (c) have more strongly-coupled initial conditions. 
}
\label{fig:test}
\end{figure}

\begin{figure}
\centering
\begin{subfigure}{.3\textwidth}
  \centering
    \caption{$\Lambda/\Lambda_0=10$}
 \includegraphics[width=.9\linewidth]{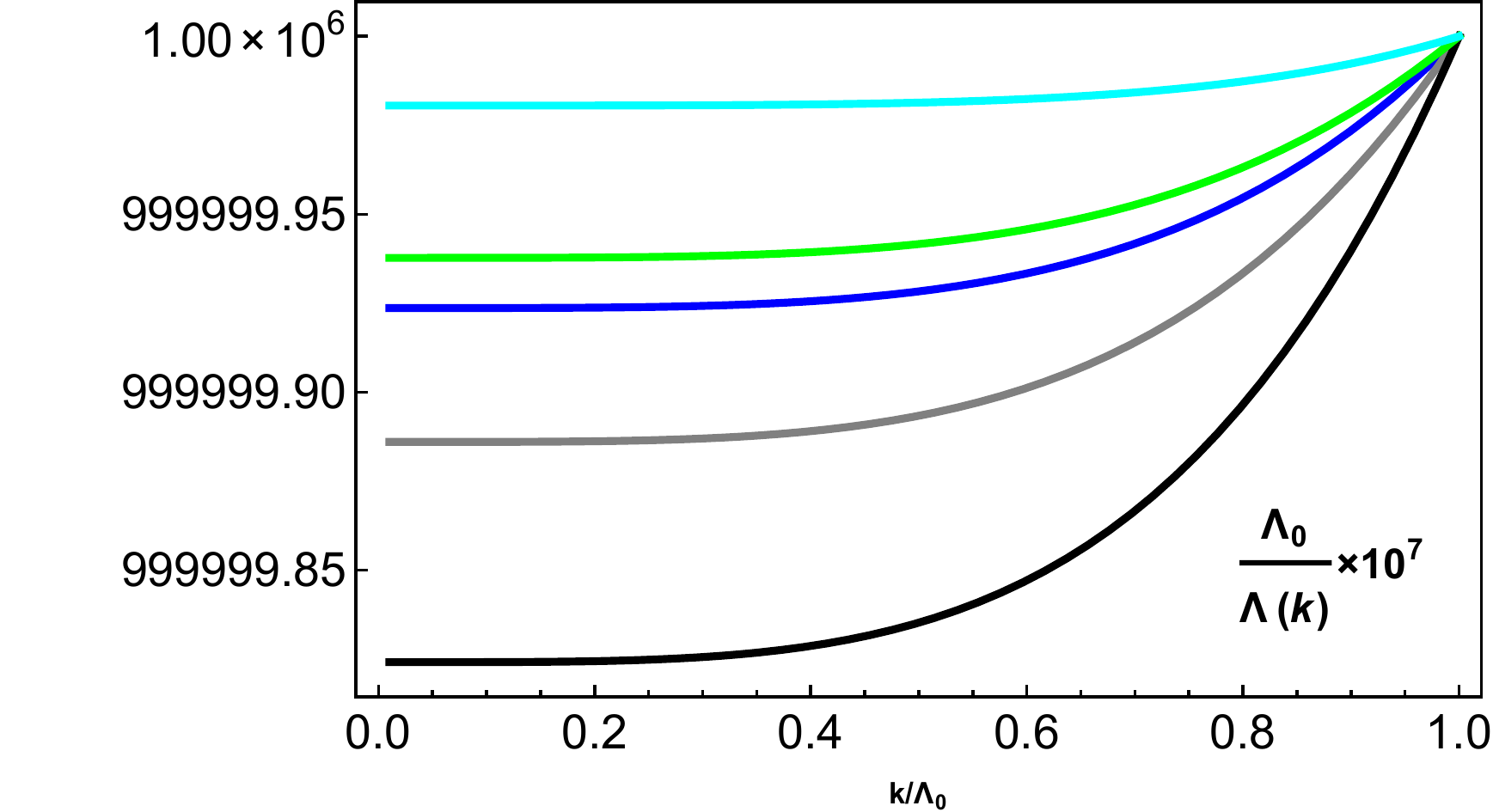}
  \label{fig:sub1sc}
\end{subfigure}
\begin{subfigure}{.3\textwidth}
  \centering
    \caption{$\Lambda/\Lambda_0 =1$}
  \includegraphics[width=.9\linewidth]{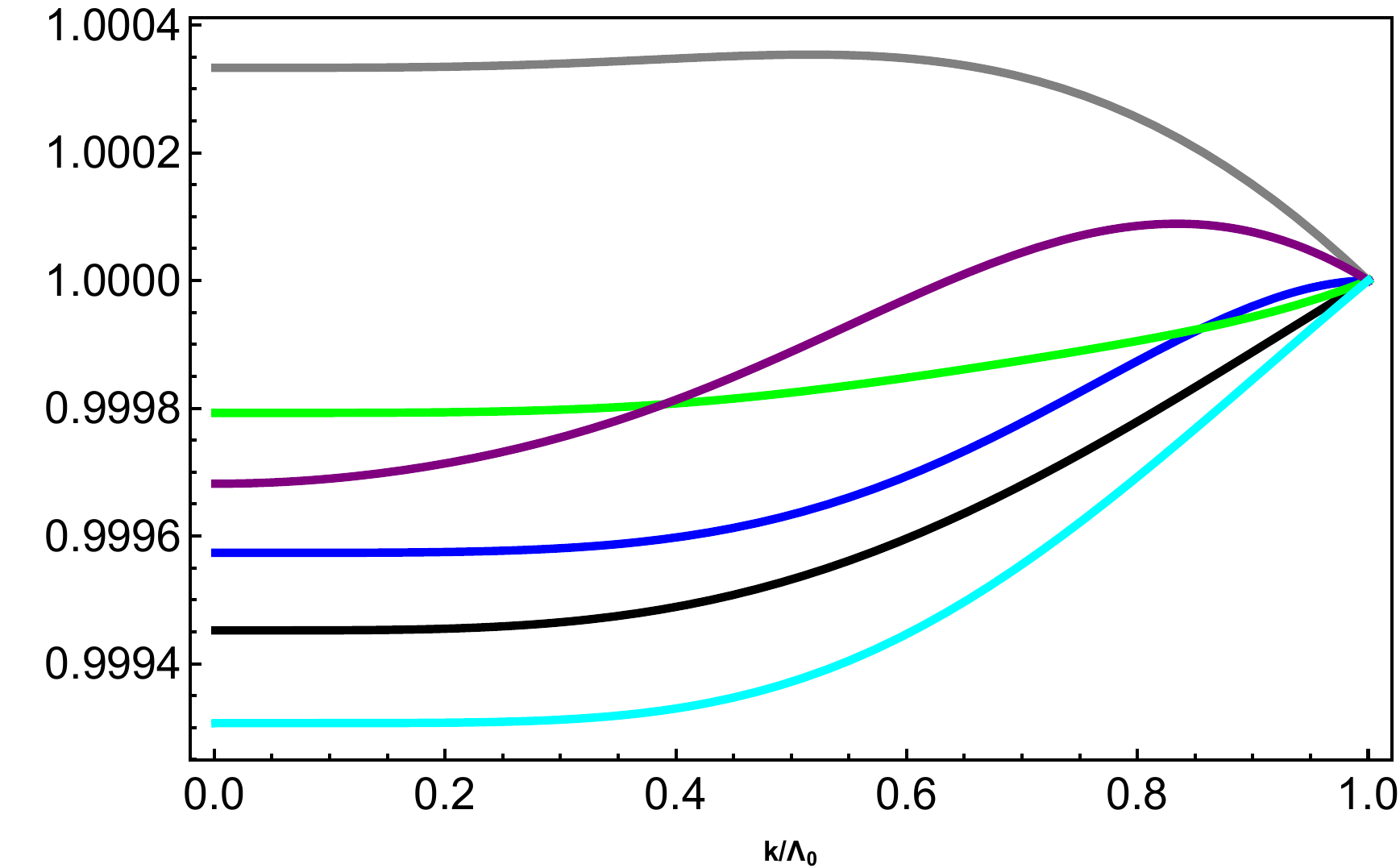}
  \label{fig:sub2sc}
\end{subfigure}
\begin{subfigure}{.3\textwidth}
  \centering
    \caption{$\Lambda/\Lambda_0=0.1$}
  \includegraphics[width=.9\linewidth]{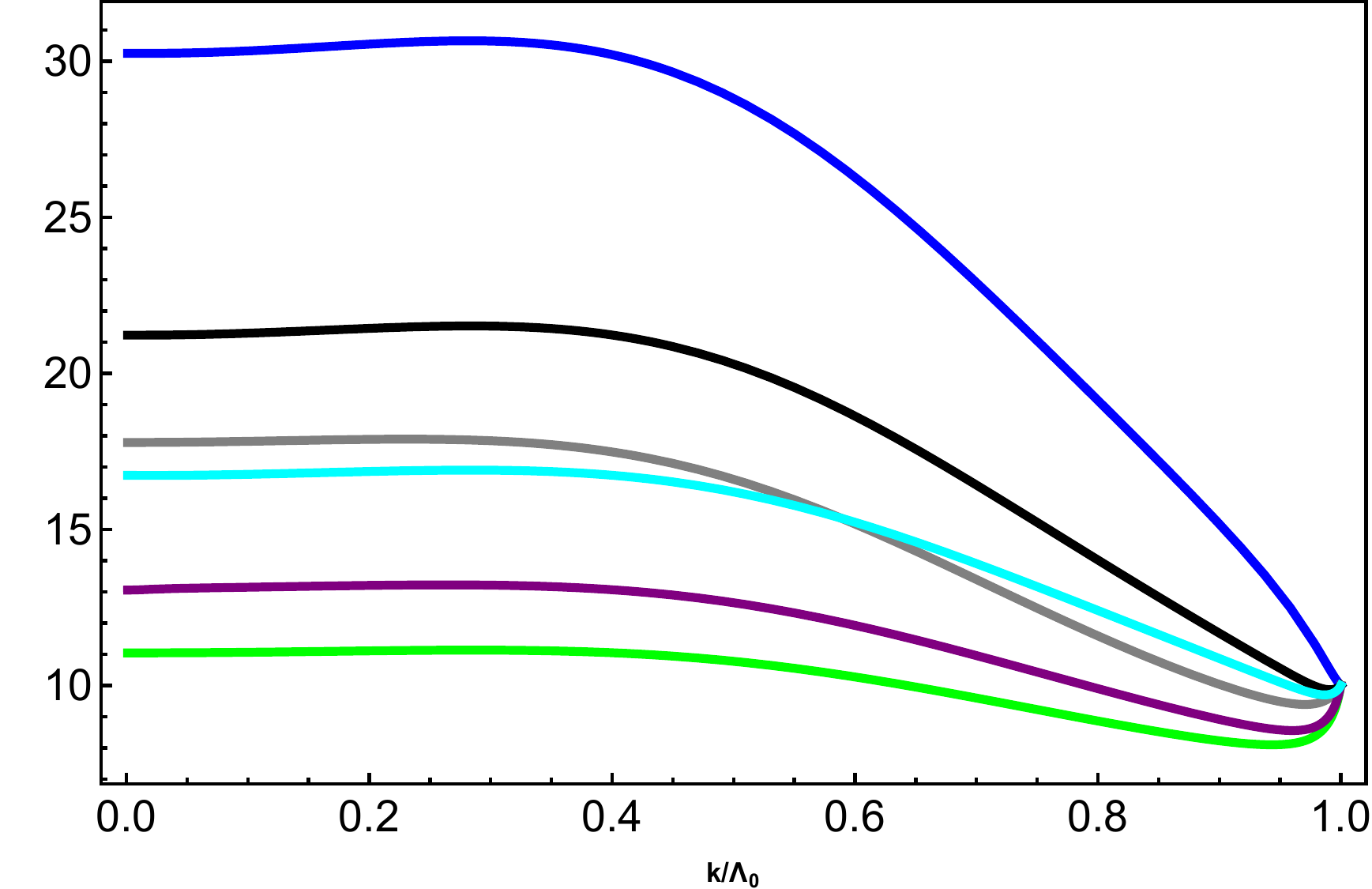}
  \label{fig:sub3sc}
\end{subfigure}%
\caption{
The running of the inverse strong-coupling scale in units of the cut-off, $\Lambda_0/\Lambda(k)$, associated with each interaction operator for the same flows as presented in FIG. \ref{fig:test}, adopting the same colour schemes.  The ``dressed'' strong-coupling length scale takes into accounting the wavefunction renormalisation of the scalar field at a given scale, and is defined as $1/\Lambda_{n,m}(k)=\left(c_{n, m}(k) / Z_1(k)^{n+\frac{m}{2}}\right)^{1/(4n+3m-4)}$.  For the  case (a) we did not include the running strong-coupling scale associated with  $c_{1,2}$ for presentation purposes, since it runs to a  slightly smaller length scale in comparison to the other ones. For the same plot, we further rescaled the curves by a factor of $10^7$ in order to properly display their very gentle running.
}
\label{fig:testsc}
\end{figure}

Despite the absence of asymptotic safety,  we can still study the RG flow for this class of derivatively coupled scalar theories, treating them as effective theories, valid only up to some finite  cut-off, $\Lambda_0$, with the boundary condition at $\Lambda_0$ set by matching to whatever UV-completes the theory.  Assuming initial conditions such that $c_{n, m}(\Lambda_0) \sim {\cal O}(1)/\Lambda^{4n+3m-4}$, we have evolved  the flows numerically down to low scales for the following three generic cases: $\Lambda_0 \ll \Lambda$,  $\Lambda_0 \sim \Lambda$ and  $\Lambda_0 \gg \Lambda$. In FIG. \ref{fig:test} we plot the running of each (dimensionful) coupling, while FIG. \ref{fig:testsc} shows the running of the ``dressed'' length scale associated with the strength of each interaction, taking into account wavefunction renormalisation.
When the bare action is weakly-coupled $\Lambda> \Lambda_0$, the couplings generically run to smaller values at low energies (FIG. \ref{fig:sub1}), with the length scale of strong coupling for all interactions becoming slightly smaller (FIG. \ref{fig:sub1sc}). The running is very gentle, consistent with the fact that we are always close to the gaussian fixed point.  When the bare action is more strongly-coupled $\Lambda \lesssim \Lambda_0$, as shown in FIGs \ref{fig:sub2} and \ref{fig:sub3}, we see that there is some initial running of the couplings, but that they quickly settle down to fixed values. From FIG. \ref{fig:sub3sc} we see that this feeble amount of running kicks in whilst we are still in the strongly-coupled phase. Of course, the polynomial truncation described above by \eqref{trunc} implicitly assumes a perturbative background with gentle gradients and is not particularly well suited to the weakly-coupled regime. Nevertheless, as we will see shortly by studying non-perturbative backgrounds directly, this feeble amount of running at strong coupling is a generic feature  and is easily understood in terms of a quantum Vainshtein mechanism.  From FIG. \ref{fig:sub3sc} we also observe that there is an initial increase in the length scale of strong coupling of each interaction, such that the infra-red theory sees interactions suppressed by a lower energy scale in comparison to the bare theory.  We have seen that this behaviour is actually quite generic and in contrast to the special case studied in \cite{riding}\footnote{For the case of $P(X)=Z_1X+c_{0,2} X^2$, \cite{riding} showed that the length scale of strong coupling is shorter in the infra-red theory. We also see this behaviour when we choose that particular boundary condition, but it is not generic for all boundary conditions. We attribute this effect to the higher-order interactions not included in \cite{riding}. Considering the flow from low to high energies we would expect those higher-order interactions to renormalise the wavefunction to larger values in the UV.
}.

We  conclude this section by asking what happens in the {\it extremely non-perturbative regime},  i.e. on backgrounds for which $|X | \gg \Lambda^4, ~|B| \gg  \Lambda^3$, relevant for Vainshtein screening. We should not be concerned that solutions exhibiting Vainshtein screening around heavy sources do not correlate directly to our ansatz for $\phi$ (with constant $B$). What is qualitatively important about Vainshtein screening is the dominance of higher order operators in the presence of  large gradients, such that we obtain large $Z$ factors, and our ansatz allows us to mimic these qualitative features. Further, whenever Vainshtein screening is operative there is often some particular interaction that dominates the strongly-coupled background, let us assume that this is indeed so and that the background is dominated by one particular polynomial interaction\footnote{It would be straightforward to generalise this argument to various non-polynomial interactions.}, $c_{n, m} X^n B^m$, where $4n+3m-4>2$ and $c_{n, m} (\Lambda_0) \sim {\cal O}(1)/\Lambda^{4n+3m-4}$.  We therefore approximate the running  Lagrangian as $$P_k(X, B) \approx Z_1(k) X+c_{n, m}(k)X^n B^m \ ,$$ assumed to be valid when   $|X|  \gg \Lambda^4, ~|B| \gg  \Lambda^3$, or equivalently, as $\Lambda \to 0$. Plugging this ansatz into the ERG equation \eqref{erg1} and using the fact that $X/B^2 =-x^2/32$, we obtain 
\be
\del_t P_k= \frac{1}{(2\pi)^3}  \int_0^k dp p^3 \int_0^\pi d\theta \sin^2\theta \frac{\del_t(-Z_1 k^2)-p^2\del_t Z_1}{-Z_1 k^2+c_{n, m}X^{n-2} B^{m+2} p_\theta ( xp) } ,
 \ee
 where $p_{\theta}(z)$ is a fourth-order polynomial with angular dependent coefficients. Given its boundary behaviour, we express the coupling as $c_{n, m} (k) \sim d_{n, m}(k)/\Lambda^{4n+3m-4}$ with the assumption that $d_{n, m}(k) \sim {\cal O}(1)$.  This is certainly true near the boundary and will prove to be a self-consistent assumption on solving for the resulting flow. Indeed, as $\Lambda \to 0$, we now find that the integrand on the right-hand side of this equation is generically suppressed. In principle, this suppression could be absent thanks to cancellations in $p_\theta ( xp)$. However, these cancellations will only occur on contributions to the integral  of measure zero and will only matter  if the integrand also diverges  at precisely those points. As this is clearly not the case we can safely infer that the integral itself is suppressed. 
 
 The suppression of the integral as $\Lambda \to 0$ implies an absence of running. A similar conclusion was drawn in \cite{riding} and the result is very easy to understand thanks to a comparison with   Vainshtein screening \citevain. Classically, the Vainshtein mechanism works when the background induces a large {\it Z factor} for  the classical fluctuations. For a single scalar, the fluctuations about the strong background go as $\delta {\cal L} \supset Z^{\mu\nu}[\phi] \del_\mu \delta \phi \del _\nu \delta \phi $ and when $Z^{\mu\nu}[\phi]$ is large, the classical fluctuations begin to decouple from external sources. Schematically, in the above example, we have that  $Z \sim \left|\frac{d_{n, m}}{\Lambda^{4n+3m-4} }  X^{n-1} B^m \right| \gg 1$ in the limit $|X| \gg \Lambda^4, ~|B| \gg  \Lambda^3$, given that  the $d_{m,n}$  remain order one. What we are seeing in the absence of running  is a quantum version of Vainshtein screening in a non-perturbative setting.  On these strong backgrounds, the quantum fluctuations begin to decouple and the RG running is suppressed.

  The absence of running also suggests that we have a non-trivial  fixed point in the limit $\Lambda \to 0$. However, this is not asymptotic safety since the dimensionless ratio of the coupling to either a fixed  scale, or the RG scale,  is divergent. Indeed, what is unusual about the $\Lambda \to 0$ limit  is that it is {\it infinitely} strongly-coupled.  As a result,  the dynamical dimension of the scalar is no longer given by its engineering dimension, but by $d_\phi=\frac{4-2(n+m)}{2n+m}$.  Because $c_{m, n}$ and, equivalently, $d_{m, n}$ do not run, the leading-order part of the action, $\int d^4 x \frac{d_{n, m}}{\Lambda^{4n+3m-4}}  X^n B^m$ remains invariant under a rescaling $x^\mu \to  \lambda x^\mu, \phi \to  \lambda^{-d_\phi} \phi$, consistent with a scale-invariant fixed point. Writing $\Lambda=\epsilon \Lambda_*$ where $\Lambda_*$ is some fixed finite scale and $\epsilon$ is small, we can absorb the powers of $\epsilon$ into a redefinition of $\hbar$ and understand the $\Lambda \to 0$ limit as a {\it classical} scale-invariant limit.  This is entirely consistent with the Wilsonian perspective on classicalisation \cite{class} advocated in \cite{Kovner} (see also discussion in \cite{qdgp}) for which the lack of running could have been regarded as a smoking gun \cite{Vik,Tet4}.  Of course, UV completion through classicalisation is only possible if there exist  static self-sourcing classicalon profiles whose radius grows with energy. Whether or not they exist can, in turn, be sensitive to the signs of couplings constants in the classical action \cite{roadsign}. Clearly our background ansatz does not correspond to a self-sourced classicalon solution, but by assuming that a non-linear interaction is dominating the dynamics,  we are endowing the background with one of  the classicalon's most important features. What we have shown here is the off--shell impact of this through the suppression of running, staying frozen to the   UV initial condition in  the absence of relevant deformations\footnote{Deformations about the infinitely strongly-coupled fixed point are classified using the scaling dimension of the field which, as we have seen,  differs from its engineering dimension.  One might hope to treat these deformations perturbatively using the dual formulations outlined in \cite{classdual1, classdual2}.} that could drive the RG flow to the infrared. 
  
Although classicalisation is an interesting alternative to the standard Wilsonian path to a UV completion, we might wish to avoid it in the context of large-distance  modifications of gravity, with Vainshtein screening, to avoid running into the potential issues  outlined in the first part of this paper.  Adopting a more conventional approach,  we could consider the implications of this behaviour within the context of a standard Wilsonian framework, with an effective field theory cut off at some finite scale. Then the lack of running  means that the low-energy dynamics is highly sensitive to the boundary condition at the UV cut-off and therefore the details of UV matching.  The predictive power of effective field theory is lost in such a scenario. To proceed, one probably has to work with a top down approach and hope for a conventional UV-complete model in which the large-scale observables are  screened close to a source on account of some non-perturbative mixing between the light and heavy sectors of the theory.  

 \section{Summary}

In the first part of this paper, we have critically assessed the status of Vainshtein screening from an ultra-violet perspective. By its very nature, Vainshtein screening probes strongly-coupled dynamics of the underlying theory on observable scales. This inevitably pushes the UV sector of theory out to macroscopic scales, rendering  an understanding of the UV properties is essential if we are to make any trustworthy predictions. The least radical scenario would be one in which new degrees of freedom help preserve perturbative unitarity, yet do not spoil the effects of screening.  This would correspond to a UV-complete theory of Vainshtein screening, in which the heavy sector plays a crucial role and cannot be ignored. In the absence of new degrees of freedom, one is faced with two further possibilities.  One of these is UV completion through asymptotic safety \cite{Percacci, Rosten, Niedermaier}. This is investigated in the second part of our paper as part of a broader analysis of the (non-perturbative) RG flow  in derivatively coupled scalar theories. We find no evidence for a non-gaussian fixed point consistent with asymptotic safety. 

Staying within a conventional Wilsonian framework, where the derivatively coupled scalar theory is treated as an effective field theory valid up to some cut-off, we studied the corresponding RG flow showing how on a weakly-coupled perturbative background the theory runs towards the (trivial) gaussian fixed point in the infra-red. This means that the interactions in the low-energy theory are suppressed by larger scales than their UV counterpart, in accordance with their irrelevant nature.  On a strongly-coupled background, we find that running is suppressed. This is consistent with the Wilsonian perspective on classicalisation \cite{class} alluded to in \cite{Kovner} and is the second alternative to integrating in new degrees of freedom. Although classicalisation may be an interesting way to UV-complete this type of theory, we have argued that its application to large-scale modified gravity scenarios is  potentially problematic for collider experiments because  the classicalisation scale is so low. The details of this certainly merit further investigation.  

Finally, let us compare aspects of our work with that of  \cite{riding}. Our analysis generalises their results to a wider class of models finding schematically similar behaviour in both the weakly-coupled and strongly-coupled, non-perturbative regimes. However, our interpretation of the latter is somewhat different.  Although we agree that there is no running in this regime, we argue that this can be understood as a scale-invariant UV fixed point exhibiting classical behaviour. To properly understand its quantum deformations we need an appropriate perturbative treatment, possibly adopting the dual formalisms proposed in  \cite{classdual1, classdual2}. 


\section*{Acknowledgements}

We would  like to thank Clare Burrage, John Butterworth, Peter Millington, Florian Niedermann,  Paul Saffin, Ignacy Sawicki, David Stefanyszyn and Alex Vikman for useful discussions and feedback. AP is funded in part by Leverhulme Research Project Grant, and IDS is supported by ESIF and MEYS (Project CoGraDS -- CZ.02.1.01/0.0/0.0/15\_003/0000437).  AP would like to encourage anyone reading this paper to donate to  \url{https://www.justgiving.com/crowdfunding/help4matt} in support of a close friend.


\appendix

\section{Absence of a non-gaussian fixed point for $P(X)$.}
We wish to show  the absence of a non-trivial fixed-point solution of the ERG equation \eqref{erg1} for a generic function $P(X)$, not necessarily of a polynomial form. To this end  we  assume $P(X, B) \to P(X)$ in  the  ERG equation \eqref{erg1} so that it becomes
 \be \label{erg2}
 \del_t P_k(X)=\frac12 \int_{|p| \leq k} \frac{d^4 p}{(2\pi)^4} \frac{-\del_t(Z_1 k^2)-p^2\del_t Z_1}{\alpha_0 + i\alpha_1 x_\mu p^\mu+\alpha_2  ( x_\mu p^\mu)^2},
 \ee
where  \be
 \alpha_0 = -Z_1 k^2+p^2\left( Z_1-P_X \right), \qquad
 \alpha_1 = -\frac{B^2}{8} \left( 3 P_{XX}  +XP_{X^3} \right), \qquad
 \alpha_2 = \frac{B^2}{16} P_{XX},
 \ee
 and we remind the reader that we work with the background choice  $\phi = \frac{B}{8}x_{\mu}x^{\mu}$, and $B = \Box \phi = \text{constant}$ (see discussion around equation (\ref{erg1})).  Now, as there is no $B$-dependence on the left-hand side of \eqref{erg2}, we require that the same should be true on the right, for the equation to close under the $P(X)$ form. In general, there is $B$-dependence in $\alpha_{1}$ and $\alpha_2$, as well as in $x^\mu$.  However, we recall that for the choice of background ansatz we can always trade $B^2 |x|^2 \to -32 X$, allowing us to eliminate the $B$-dependence in the combination $\alpha_2 ( x_\mu p^\mu)^2$, but not in $i\alpha_1 x_\mu p^\mu$. It follows that $\alpha_1$ must be set to zero to have any hope of identifying a fixed point. Setting $\alpha_1=0$ leads to a differential equation for $P(X)$ which is easily solved at the fixed point to give
\be \label{sol}
P_k(X)=\tilde V k^4 +Z_1 X+\frac{\tilde c_{-1}}{X} k^8,
\ee
where $\tilde V, Z_1$ and $\tilde c_{-1}$ are all constants on the RG flow. The last term in this expression could easily be eliminated on grounds of locality -- even if we allow for it in principle, we find that it is forced to be absent at a fixed point anyway. To see this, we now plug the solution \eqref{sol} back into the ERG equation \eqref{erg2}, and perform the angular integration as in the main text to obtain, 
\be
4 \tilde V  +8 \frac{\tilde c_{-1}}{\tilde X}  =\frac{4 \pi}{(2\pi)^4}\int_0^1 d\tilde p \tilde p^3 \int_0^\pi d\theta \sin^2\theta \frac{1}{1+ \frac{ \tilde c_{-1}  \tilde  p^2}{Z_1 \tilde X^2} (4 \cos^2\theta-1)},
\ee
with $\tilde X=X/k^4$ and $\tilde p=p/k$. This equation should hold for all $\tilde X$. However, it is easy to check that this can only be true when $\tilde c_{-1}=0$ and $\tilde V=1/128 \pi^2$. This corresponds to the gaussian fixed point, or in other words, the only fixed-point solution is a free theory. We have thus proved the absence of the non-gaussian fixed point within the $P(X)$ class. We close by reminding the reader that the existence of a non-trivial fixed point should be evident for any background choice, while the opposite is not in principle true.


\begin{thebibliography}{0}%
\makeatletter
\providecommand \@ifxundefined [1]{%
 \@ifx{#1\undefined}
}%
\providecommand \@ifnum [1]{%
 \ifnum #1\expandafter \@firstoftwo
 \else \expandafter \@secondoftwo
 \fi
}%
\providecommand \@ifx [1]{%
 \ifx #1\expandafter \@firstoftwo
 \else \expandafter \@secondoftwo
 \fi
}%
\providecommand \natexlab [1]{#1}%
\providecommand \enquote  [1]{``#1''}%
\providecommand \bibnamefont  [1]{#1}%
\providecommand \bibfnamefont [1]{#1}%
\providecommand \citenamefont [1]{#1}%
\providecommand \href@noop [0]{\@secondoftwo}%
\providecommand \href [0]{\begingroup \@sanitize@url \@href}%
\providecommand \@href[1]{\@@startlink{#1}\@@href}%
\providecommand \@@href[1]{\endgroup#1\@@endlink}%
\providecommand \@sanitize@url [0]{\catcode `\\12\catcode `\$12\catcode
  `\&12\catcode `\#12\catcode `\^12\catcode `\_12\catcode `\%12\relax}%
\providecommand \@@startlink[1]{}%
\providecommand \@@endlink[0]{}%
\providecommand \url  [0]{\begingroup\@sanitize@url \@url }%
\providecommand \@url [1]{\endgroup\@href {#1}{\urlprefix }}%
\providecommand \urlprefix  [0]{URL }%
\providecommand \Eprint [0]{\href }%
\providecommand \doibase [0]{http://dx.doi.org/}%
\providecommand \selectlanguage [0]{\@gobble}%
\providecommand \bibinfo  [0]{\@secondoftwo}%
\providecommand \bibfield  [0]{\@secondoftwo}%
\providecommand \translation [1]{[#1]}%
\providecommand \BibitemOpen [0]{}%
\providecommand \bibitemStop [0]{}%
\providecommand \bibitemNoStop [0]{.\EOS\space}%
\providecommand \EOS [0]{\spacefactor3000\relax}%
\providecommand \BibitemShut  [1]{\csname bibitem#1\endcsname}%
\let\auto@bib@innerbib\@empty
\end{thebibliography}%


\begin{thebibliography}{10}
 

 
 
\bibitem{kinf}
  C.~Armendariz-Picon, T.~Damour and V.~F.~Mukhanov,
  Phys.\ Lett.\ B {\bf 458} (1999) 209
  [hep-th/9904075].
  
\bibitem{kess}
  T.~Chiba, T.~Okabe and M.~Yamaguchi,
  Phys.\ Rev.\ D {\bf 62} (2000) 023511
  [astro-ph/9912463].
  
\bibitem{edreview}
  E.~J.~Copeland, M.~Sami and S.~Tsujikawa,
  Int.\ J.\ Mod.\ Phys.\ D {\bf 15} (2006) 1753
  [hep-th/0603057].
  
\bibitem{myreview}
  T.~Clifton, P.~G.~Ferreira, A.~Padilla and C.~Skordis,
  Phys.\ Rept.\  {\bf 513} (2012) 1
  [arXiv:1106.2476 [astro-ph.CO]].
  
\bibitem{justinreview}
  A.~Joyce, B.~Jain, J.~Khoury and M.~Trodden,
  Phys.\ Rept.\  {\bf 568} (2015) 1
  [arXiv:1407.0059 [astro-ph.CO]].
  
\bibitem{Vainshtein}
  A.~I.~Vainshtein,
  Phys.\ Lett.\  {\bf 39B} (1972) 393.
  
\bibitem{vainintro}
  E.~Babichev and C.~Deffayet,
  Class.\ Quant.\ Grav.\  {\bf 30} (2013) 184001
  [arXiv:1304.7240 [gr-qc]].
  
\bibitem{dyson}
  N.~Kaloper, A.~Padilla and N.~Tanahashi,
  JHEP {\bf 1110} (2011) 148
  [arXiv:1106.4827 [hep-th]].
  
\bibitem{cham1}
  J.~Khoury and A.~Weltman,
  Phys.\ Rev.\ Lett.\  {\bf 93} (2004) 171104
  doi:10.1103/PhysRevLett.93.171104
  [astro-ph/0309300].
  
\bibitem{cham2}
  J.~Khoury and A.~Weltman,
  Phys.\ Rev.\ D {\bf 69} (2004) 044026
  [astro-ph/0309411].
  
\bibitem{symm1}
  K.~Hinterbichler and J.~Khoury,
  Phys.\ Rev.\ Lett.\  {\bf 104} (2010) 231301
  [arXiv:1001.4525 [hep-th]].
  
\bibitem{symm2}
  K.~Hinterbichler, J.~Khoury, A.~Levy and A.~Matas,
  Phys.\ Rev.\ D {\bf 84} (2011) 103521
  [arXiv:1107.2112 [astro-ph.CO]].
  
\bibitem{will}
  C.~M.~Will,
  Living Rev.\ Rel.\  {\bf 17} (2014) 4
  [arXiv:1403.7377 [gr-qc]].
  
\bibitem{florian}
  F.~Niedermann and A.~Padilla,
  arXiv:1706.04778 [hep-th].

\bibitem{wein}
  S.~Weinberg,
  Rev.\ Mod.\ Phys.\  {\bf 61} (1989) 1.
\bibitem{cliff}
  C.~P.~Burgess,
  arXiv:1309.4133 [hep-th].
\bibitem{me}
  A.~Padilla,
  arXiv:1502.05296 [hep-th].
  
\bibitem{power}
  G.~Dvali,
  New J.\ Phys.\  {\bf 8} (2006) 326
  [hep-th/0610013].
  
\bibitem{unit}
  N.~Kaloper, A.~Padilla, P.~Saffin and D.~Stefanyszyn,
  Phys.\ Rev.\ D {\bf 91} (2015) no.4,  045017
  [arXiv:1409.3243 [hep-th]].
  
  
\bibitem{nk}
  G.~D'Amico, N.~Kaloper and A.~Lawrence,
  arXiv:1709.07014 [hep-th].
  
\bibitem{georgi}
  A.~Manohar and H.~Georgi,
  Nucl.\ Phys.\ B {\bf 234} (1984) 189.

\bibitem{manohar}
  B.~M.~Gavela, E.~E.~Jenkins, A.~V.~Manohar and L.~Merlo,
  Eur.\ Phys.\ J.\ C {\bf 76} (2016) no.9,  485
  [arXiv:1601.07551 [hep-ph]].
  
\bibitem{GW170817}
  B.~P.~Abbott {\it et al.} [LIGO Scientific and Virgo Collaborations],
  Phys.\ Rev.\ Lett.\  {\bf 119} (2017) no.16,  161101
  [arXiv:1710.05832 [gr-qc]].
  
\bibitem{GRB1}
  A.~Goldstein {\it et al.},
  Astrophys.\ J.\  {\bf 848} (2017) no.2,  L14
  [arXiv:1710.05446 [astro-ph.HE]].
  
\bibitem{GRB2}
  V.~Savchenko {\it et al.},
  Astrophys.\ J.\  {\bf 848} (2017) no.2,  L15
  [arXiv:1710.05449 [astro-ph.HE]].
  
\bibitem{GRB3}
  B.~P.~Abbott {\it et al.} [LIGO Scientific and Virgo and Fermi-GBM and INTEGRAL Collaborations],
  Astrophys.\ J.\  {\bf 848} (2017) no.2,  L13
  [arXiv:1710.05834 [astro-ph.HE]].
  
\bibitem{GRB4}
  [LIGO Scientific and Virgo and Fermi GBM and INTEGRAL and IceCube and IPN and Insight-Hxmt and ANTARES and Swift and Dark Energy Camera GW-EM and DES and DLT40 and GRAWITA and Fermi Large Area Telescope and ATCA and ASKAP and OzGrav and DWF (Deeper Wider Faster Program) and AST3 and CAASTRO and VINROUGE and MASTER and J-GEM and GROWTH and JAGWAR and CaltechNRAO and TTU-NRAO and NuSTAR and Pan-STARRS and KU and Nordic Optical Telescope and ePESSTO and GROND and Texas Tech University and TOROS and BOOTES and MWA and CALET and IKI-GW Follow-up and H.E.S.S. and LOFAR and LWA and HAWC and Pierre Auger and ALMA and Pi of Sky and DFN and ATLAS and High Time Resolution Universe Survey and RIMAS and RATIR and SKA South Africa/MeerKAT Collaborations and AstroSat Cadmium Zinc Telluride Imager Team and AGILE Team and 1M2H Team and Las Cumbres Observatory Group and MAXI Team and TZAC Consortium and SALT Group and Euro VLBI Team and Chandra Team at McGill University],
  Astrophys.\ J.\  {\bf 848} (2017) L12
  [arXiv:1710.05833 [astro-ph.HE]].
  
  
\bibitem{Sak}
  J.~Sakstein and B.~Jain,
  arXiv:1710.05893 [astro-ph.CO].
  
\bibitem{Spanish}
  J.~M.~Ezquiaga and M.~Zumalacrregui,
  arXiv:1710.05901 [astro-ph.CO].
  
\bibitem{vern}
  P.~Creminelli and F.~Vernizzi,
  arXiv:1710.05877 [astro-ph.CO].
  
\bibitem{pedro}
  T.~Baker, E.~Bellini, P.~G.~Ferreira, M.~Lagos, J.~Noller and I.~Sawicki,
  arXiv:1710.06394 [astro-ph.CO].
  
\bibitem{Amendola}
  L.~Amendola, M.~Kunz, I.~D.~Saltas and I.~Sawicki,
  arXiv:1711.04825 [astro-ph.CO].

\bibitem{Deffayet:2010qz}
  C.~Deffayet, O.~Pujolas, I.~Sawicki and A.~Vikman,
  JCAP {\bf 1010} (2010) 026
  [arXiv:1008.0048 [hep-th]].
  

\bibitem{piazza}
  J.~Beltran Jimenez, F.~Piazza and H.~Velten,
  Phys.\ Rev.\ Lett.\  {\bf 116} (2016) no.6,  061101
  [arXiv:1507.05047 [gr-qc]].
  
\bibitem{karim}
  D.~Langlois, R.~Saito, D.~Yamauchi and K.~Noui,
  arXiv:1711.07403 [gr-qc].
  
\bibitem{marco}
  M.~Crisostomi and K.~Koyama,
  arXiv:1711.06661 [astro-ph.CO].

\bibitem{cedric}
  E.~Babichev, C.~Deffayet and G.~Esposito-Farese,
  Phys.\ Rev.\ Lett.\  {\bf 107} (2011) 251102
  [arXiv:1107.1569 [gr-qc]].

  
\bibitem{gal}
  A.~Nicolis, R.~Rattazzi and E.~Trincherini,
  Phys.\ Rev.\ D {\bf 79} (2009) 064036
  [arXiv:0811.2197 [hep-th]].
  
\bibitem{nogal}
  J.~Renk, M.~Zumalacrregui, F.~Montanari and A.~Barreira,
  JCAP {\bf 1710} (2017) 020
  [arXiv:1707.02263 [astro-ph.CO]].
  
\bibitem{nima}
  A.~Adams, N.~Arkani-Hamed, S.~Dubovsky, A.~Nicolis and R.~Rattazzi,
  JHEP {\bf 0610} (2006) 014
  [hep-th/0602178].
  
  
\bibitem{class}
  G.~Dvali, G.~F.~Giudice, C.~Gomez and A.~Kehagias,
  JHEP {\bf 1108} (2011) 108
  [arXiv:1010.1415 [hep-ph]].
  
\bibitem{Kovner}
  A.~Kovner and M.~Lublinsky,
  JHEP {\bf 1211} (2012) 030
  [arXiv:1207.5037 [hep-th]].
  
  
\bibitem{UVgal}
  L.~Keltner and A.~J.~Tolley,
  arXiv:1502.05706 [hep-th].
  
\bibitem{apples}
  P.~Millington, F.~Niedermann and A.~Padilla,
  arXiv:1707.06931 [hep-th].
  
\bibitem{energys}
  A.~Nicolis, R.~Rattazzi and E.~Trincherini,
  JHEP {\bf 1005} (2010) 095
   Erratum: [JHEP {\bf 1111} (2011) 128]
  [arXiv:0912.4258 [hep-th]].
  
\bibitem{Bellazzini}
  B.~Bellazzini, F.~Riva, J.~Serra and F.~Sgarlata,
  arXiv:1710.02539 [hep-th].
  
  \bibitem{Y3project}
  R.~Daniel and J.~Puhalo-Smith,
  University of Nottingham Physics Project (F33PJA)
  
\bibitem{pos}
  C.~de Rham, S.~Melville and A.~J.~Tolley,
  arXiv:1710.09611 [hep-th].
  
  
\bibitem{dbiV}
  C.~Burrage and J.~Khoury,
  Phys.\ Rev.\ D {\bf 90} (2014) no.2,  024001
  [arXiv:1403.6120 [hep-th]].
  
\bibitem{kick}
  A.~Padilla, E.~Platts, D.~Stefanyszyn, A.~Walters, A.~Weltman and T.~Wilson,
  JCAP {\bf 1603} (2016) no.03,  058
  [arXiv:1511.05761 [hep-th]].
  
  
\bibitem{gia}
  G.~Dvali,
  Subnucl.\ Ser.\  {\bf 53} (2017) 189
  [arXiv:1607.07422 [hep-th]].
  
\bibitem{mg}
  N.~Arkani-Hamed, H.~Georgi and M.~D.~Schwartz,
  Annals Phys.\  {\bf 305} (2003) 96
  [hep-th/0210184].
  
  
\bibitem{groj}
  C.~Grojean and R.~S.~Gupta,
  JHEP {\bf 1205} (2012) 114
  [arXiv:1110.5317 [hep-ph]].
  
\bibitem{Higgspl1}
  V.~V.~Khoze and M.~Spannowsky,
  Nucl.\ Phys.\ B {\bf 926} (2018) 95
  [arXiv:1704.03447 [hep-ph]].
  
\bibitem{Higgspl2}
  V.~V.~Khoze,
  JHEP {\bf 1706} (2017) 148
  [arXiv:1705.04365 [hep-ph]].
  
\bibitem{Higgspl3}
  V.~V.~Khoze and M.~Spannowsky,
  Phys.\ Rev.\ D {\bf 96} (2017) no.7,  075042
  [arXiv:1707.01531 [hep-ph]].
  
\bibitem{Higgsplprec}
  V.~V.~Khoze, J.~Reiness, M.~Spannowsky and P.~Waite,
  arXiv:1709.08655 [hep-ph].
  
\bibitem{PX}
  E.~Babichev, C.~Deffayet and R.~Ziour,
  Int.\ J.\ Mod.\ Phys.\ D {\bf 18} (2009) 2147
  [arXiv:0905.2943 [hep-th]].
  
%
\bibitem{Bagnuls}
  C.~Bagnuls and C.~Bervillier,
  Phys.\ Rept.\  {\bf 348} (2001) 91
  [hep-th/0002034].
  
\bibitem{Berges}
  J.~Berges, N.~Tetradis and C.~Wetterich,
  Phys.\ Rept.\  {\bf 363} (2002) 223
  [hep-ph/0005122].
  
\bibitem{Gies}
  H.~Gies,
  Lect.\ Notes Phys.\  {\bf 852} (2012) 287
  [hep-ph/0611146].
  
\bibitem{Polonyi}
  J.~Polonyi,
  Central Eur.\ J.\ Phys.\  {\bf 1} (2003) 1
  [hep-th/0110026].
  
\bibitem{Delamotte}
  B.~Delamotte,
  Lect.\ Notes Phys.\  {\bf 852} (2012) 49
  [cond-mat/0702365 [cond-mat.stat-mech]].
  
  
\bibitem{Rosten}
  O.~J.~Rosten,
  Phys.\ Rept.\  {\bf 511} (2012) 177
  [arXiv:1003.1366 [hep-th]].
  
   
\bibitem{Niedermaier}
  M.~Niedermaier and M.~Reuter,
  Living Rev.\ Rel.\  {\bf 9} (2006) 5.
  
  
\bibitem{Percacci}
  R.~Percacci,
  In *Oriti, D. (ed.): Approaches to quantum gravity* 111-128
  [arXiv:0709.3851 [hep-th]].

  
\bibitem{riding}
  C.~de Rham and R.~H.~Ribeiro,
  JCAP {\bf 1411} (2014) no.11,  016
  [arXiv:1405.5213 [hep-th]].
  
\bibitem{Shap}
  T.~de Paula Netto and I.~L.~Shapiro,
  Phys.\ Lett.\ B {\bf 716} (2012) 454
  [arXiv:1207.0534 [hep-th]].
  
\bibitem{Tet1}
  N.~Brouzakis, A.~Codello, N.~Tetradis and O.~Zanusso,
  Phys.\ Rev.\ D {\bf 89} (2014) no.12,  125017
  [arXiv:1310.0187 [hep-th]].
  
\bibitem{Tet2}
  N.~Brouzakis and N.~Tetradis,
  Phys.\ Rev.\ D {\bf 89} (2014) no.12,  125004
  [arXiv:1401.2775 [hep-th]].
  
\bibitem{Tet4}
  P.~Asimakis, N.~Brouzakis, A.~Katsis and N.~Tetradis,
  Phys.\ Lett.\ B {\bf 743} (2015) 75
  [arXiv:1412.4275 [hep-th]].
  
\bibitem{Tet3}
  A.~Codello, N.~Tetradis and O.~Zanusso,
  JHEP {\bf 1304} (2013) 036
  [arXiv:1212.4073 [hep-th]].
  
  \bibitem{Morris:1997xj}
  T.~R.~Morris and M.~D.~Turner,
  ``Derivative expansion of the renormalization group in O(N) scalar field theory,''
  Nucl.\ Phys.\ B {\bf 509} (1998) 637
  [hep-th/9704202].
  
\bibitem{Safari:2017tgs}
  M.~Safari and G.~P.~Vacca,
  ``Uncovering novel phase structures in $\Box ^k$ scalar theories with the renormalization group,''
  Eur.\ Phys.\ J.\ C {\bf 78} (2018) no.3,  251
  [arXiv:1711.08685 [hep-th]].
  

  
\bibitem{Vik}
  A.~Vikman,
  EPL {\bf 101} (2013) no.3,  34001
  [arXiv:1208.3647 [hep-th]].
  
\bibitem{Brax}
  P.~Brax and P.~Valageas,
  Phys.\ Rev.\ D {\bf 94} (2016) no.4,  043529
  [arXiv:1607.01129 [astro-ph.CO]].
  
\bibitem{Kurt1}
  K.~Hinterbichler, M.~Trodden and D.~Wesley,
  Phys.\ Rev.\ D {\bf 82} (2010) 124018
  [arXiv:1008.1305 [hep-th]].
  
\bibitem{Kurt2}
  G.~Goon, K.~Hinterbichler, A.~Joyce and M.~Trodden,
  JHEP {\bf 1611} (2016) 100
  [arXiv:1606.02295 [hep-th]].
  
\bibitem{Saltas1}
  I.~D.~Saltas and V.~Vitagliano,
  JCAP {\bf 1705} (2017) no.05,  020
  [arXiv:1612.08953 [hep-th]].
  

\bibitem{Saltas2}
  I.~D.~Saltas and V.~Vitagliano,
  Phys.\ Rev.\ D {\bf 95} (2017) no.10,  105002
  [arXiv:1611.07984 [hep-th]].
  
  
\bibitem{Wetterich}
  C.~Wetterich,
  Phys.\ Lett.\ B {\bf 301} (1993) 90
  [arXiv:1710.05815 [hep-th]].
  
\bibitem{Morris}
  T.~R.~Morris,
  Int.\ J.\ Mod.\ Phys.\ A {\bf 9} (1994) 2411
  [hep-ph/9308265].
  
     
  
\bibitem{Polchinski}
  J.~Polchinski,
  Nucl.\ Phys.\ B {\bf 231} (1984) 269.
  
\bibitem{call}
  C.~P.~Burgess and M.~Williams,
  JHEP {\bf 1408} (2014) 074
  [arXiv:1404.2236 [gr-qc]].
  
\bibitem{horn}
  G.~W.~Horndeski,
  Int.\ J.\ Theor.\ Phys.\  {\bf 10} (1974) 363.
  
\bibitem{dgsz}
  C.~Deffayet, X.~Gao, D.~A.~Steer and G.~Zahariade,
  Phys.\ Rev.\ D {\bf 84} (2011) 064039
  [arXiv:1103.3260 [hep-th]].
  
\bibitem{ostro1}
  M.~Ostrogradsky,
  Mem.\ Acad.\ St.\ Petersbourg {\bf 6} (1850) no.4,  385.
  
\bibitem{ostro2}
  R.~P.~Woodard,
  Scholarpedia {\bf 10} (2015) no.8,  32243
  [arXiv:1506.02210 [hep-th]].
  
\bibitem{bhorn}
  J.~Gleyzes, D.~Langlois, F.~Piazza and F.~Vernizzi,
  Phys.\ Rev.\ Lett.\  {\bf 114} (2015) no.21,  211101
  [arXiv:1404.6495 [hep-th]].
  
   \bibitem{Morris1}
  T.~R.~Morris,
  Phys.\ Lett.\ B {\bf 329} (1994) 241
  [hep-ph/9403340].
  
\bibitem{Morris2}
  T.~R.~Morris,
  Int.\ J.\ Mod.\ Phys.\ B {\bf 12} (1998) 1343
  [hep-th/9610012].

  
\bibitem{Litim}
  D.~F.~Litim,
  Phys.\ Lett.\ B {\bf 486} (2000) 92
  [hep-th/0005245].
  
\bibitem{Litim2}
  D.~F.~Litim,
  Phys.\ Rev.\ D {\bf 64} (2001) 105007
  [hep-th/0103195].
  
\bibitem{LPA}
  A.~Hasenfratz and P.~Hasenfratz,
  Nucl.\ Phys.\ B {\bf 270} (1986) 687.

\bibitem{LPAMorris}
  T.~R.~Morris,
  Phys.\ Lett.\ B {\bf 334} (1994) 355
  [hep-th/9405190].
  
\bibitem{LPAQG}
  D.~Benedetti and F.~Caravelli,
  JHEP {\bf 1206} (2012) 017
   Erratum: [JHEP {\bf 1210} (2012) 157]
  [arXiv:1204.3541 [hep-th]].
  
\bibitem{Falls:2016msz}
  K.~Falls and N.~Ohta,
  Phys.\ Rev.\ D {\bf 94} (2016) no.8,  084005
  [arXiv:1607.08460 [hep-th]].
  
  
\bibitem{qdgp}
  A.~Nicolis and R.~Rattazzi,
  JHEP {\bf 0406} (2004) 059
  [hep-th/0404159].
  
\bibitem{roadsign}
  G.~Dvali, A.~Franca and C.~Gomez,
  arXiv:1204.6388 [hep-th].
  
  
\bibitem{classdual1}
  G.~Gabadadze, K.~Hinterbichler and D.~Pirtskhalava,
  Phys.\ Rev.\ D {\bf 85} (2012) 125007
  [arXiv:1202.6364 [hep-th]].
  
\bibitem{classdual2}
  A.~Padilla and P.~M.~Saffin,
  JHEP {\bf 1207} (2012) 122
  [arXiv:1204.1352 [hep-th]].
  

  

  
  
  \end{thebibliography}
\end{document}